\documentstyle[aas2pp4,psfig]{article}

\slugcomment{Submitted 22 January 2001, Revised 15 March 2001}

\lefthead{Boyd \& Smale}
\righthead{RXTE Observations of LMC X--3}

\def\source{LMC~X--3}
\def\approxlt{\mathrel{\hbox{\rlap{\lower.55ex \hbox {$\sim$}}
        \kern-.3em \raise.4ex \hbox{$<$}}}}
\def\approxgt{\mathrel{\hbox{\rlap{\lower.55ex \hbox {$\sim$}}
        \kern-.3em \raise.4ex \hbox{$>$}}}}

\def\emphasize#1{{\sl #1\/}}
\def\kms{$\,$km$\,$s$^{-1}$}

\begin{document}

\title{X-ray and UV Orbital Phase Dependence in LMC X--3}

\author{Patricia T. Boyd \altaffilmark{1}}
\author{Alan P. Smale \altaffilmark{2}}

\affil{Laboratory for High Energy Astrophysics,
Code 662, NASA/Goddard Space Flight Center, Greenbelt, MD 20771}

\author{Joseph F. Dolan }

\affil{Laboratory for Astronomy and Solar Physics
Code 681, NASA/Goddard Space Flight Center, Greenbelt, MD 20771}

\altaffiltext{1}{Also Joint Center for Astrophysics, University of Maryland
Baltimore County}
\altaffiltext{2}{Also Universities Space Research Association}

\begin{abstract}

The black-hole binary \source\ is known to be variable on time
scales of days to years.  
We investigate X-ray and ultraviolet variability in the system as a
function of the 1.7 day binary phase using a 6.4 day observation with the 
\emphasize{Rossi X-ray Timing Explorer} (RXTE) from December 1998.
An abrupt 
14\% flux decrease, lasting nearly an entire orbit, is followed by a
return to previous flux levels.  This behavior occurs twice,
at nearly the same binary phase, but it is not present in consecutive orbits.
When the X-ray flux is
at lower intensity, a periodic amplitude modulation of 7\% is evident
in data folded modulo the orbital period.  The higher intensity
data show weaker correlation with phase.  This is the first report of X-ray
variability at the orbital period of \source.
Archival RXTE observations of LMC X--3 during a
high flux state in December 1996
show similar phase dependence.
An ultraviolet light curve obtained with the \emphasize
{High Speed Photometer}
aboard the 
\emphasize{Hubble Space Telescope} shows orbital modulation consistent
with that in the optical, caused by the ellipsoidal variation of the spatially
deformed companion.

The X-ray spectrum of \source\ can be acceptably represented by a
phenomenological
disk-black-body plus a power law.  Changes in the spectrum of \source\
during our observations are compatible with earlier observations
 during which variations in the 2-10 keV
flux are tracked closely by the disk geometry spectral model parameter.

\end{abstract}

\keywords{accretion, accretion disks --- stars: individual (LMC X--3)
--- stars: black holes --- stars: binaries: close --- X-rays: stars}

\section{Introduction}

LMC X--3, a bright  (up to 3$\times$10$^{38}$ erg~s$^{-1}$)
 black hole candidate in the Large 
Magellanic Cloud, is a highly variable X-ray source.  It exhibits
at least two distinct emission states.
During its more common 
high/soft state, the X-ray
spectrum is similar to that of other high/soft state black hole
candidates, with an "ultrasoft" component and a hard ($>$10keV)
tail.  The low/hard state, on the other hand,
 is characterized by an X-ray spectrum 
described by a pure power law $I(E)$ = $A_{pl}E^{-\Gamma}$ 
 with $\Gamma$$\sim$1.8 or less (Wilms
et al. 2000), significant
time variability with 30--50\% modulation of the intensity,
and a 0.4 Hz QPO
(Boyd et al. 2000).

The B3 V 
optical counterpart of \source\ (Warren
and Penfold 1975) (V $\sim$16.7-17.5) shows a large
velocity range (semi-amplitude K=235 \kms) through its 1.7 
day orbital period.
The lack of eclipses indicates that the
inclination of the system is $<$70$^\circ$, and this leads to a compact
object mass of $\sim$7M$_{\sun}$, making \source\ a prime black
hole candidate 
(Cowley
et al.\ 1983, Paczynski 1983, Ebisawa et al.\ 1993, but see also Mazeh
et al.\ 1986).
The optical light curve shows minima
at the times of conjunction of the two components, with 
amplitude $\sim$0.2 mag each; the approximately equal minima of the optical
light curve are consistent with an underlying ellipsoidal stellar
surface (van der Klis et al. 1985).  Comparing observations of the source
over a long time baseline, it is apparent that the ratio of optical to
X-ray flux is not constant.  This may indicate secular variation of the
disk structure over a time scale of a week, or a change in anisotropy
of the X-ray emission (Treves et al. 1990).

The long-term X-ray luminosity of the system is strongly modulated
on time scales of hundreds of days 
(Cowley et al. 1991, Wilms et al. 2000).
The mean 2-10 keV X-ray flux varies by a factor of more than one hundred
during this long-term cycle.  
This variability was previously
 attributed to the precession of a
bright, tilted, and warped
accretion disk.
The discovery of recurrent
low/hard states in \source\ argues against this mechanism being responsible 
for the
long term variation (Wilms et al. 2000).

Previous X-ray observations have detected no phase-related
variability in \source, and the 1.7-day orbital period is not detected
prominently in the RXTE ASM light curve. However, the
Ginga and EXOSAT observations of this source were quite fragmentary --
seven observations with durations ranging from 1 to 24~ksec between
December 1983 and December 1984 with EXOSAT (Treves et al.\ 1988), and a
series of much shorter exposures spread over three years with Ginga
(Cowley et al.\ 1991).  Prior to the launch of RXTE,
no intensive study of \source\ sampling an
entire binary cycle in a single epoch had ever been performed. 
We thus proposed an RXTE
observation to create a large set of data covering several consecutive
orbital cycles with the minimum of interruptions. We present here the
results obtained from this observation, performed in 1998 December,
along with a reanalysis of archival data from a shorter dedicated
RXTE pointing two years earlier.  We also include previously unpublished
data from the High Speed Photometer aboard Hubble Space Telescope
(HST), which is to our knowledge
the only ultraviolet data on this source covering the 1.7 day binary orbit.

\section{Observations}

We observed \source\ with RXTE (Bradt, Rothschild, \& Swank, 1993) 
between 1998 December 08 17:17 UT --
December 15 00:01 UT, for an on-source total good time of 283 kiloseconds.
These observations are summarized in Table 1.
The data presented here
were obtained using the PCA instrument in the Standard 2 and
Good Xenon data modes, with time resolutions of 16 sec and
$<1\mu$sec respectively.  The PCA consists of five Xe proportional
counter units (PCUs), with a combined effective area of about 6500 cm$^2$
(Jahoda et al 1996).  Only three of the five PCUs were on through the
entire observation; we analyzed data only from these three detectors.

Spectra and light curves were extracted using the RXTE standard data analysis
software, FTOOLS 5.0.  PCA background subtraction was
performed using the ``L7-240'' (v19990824/0909)
faint source model. 
\source\ is reliably detected above background out to about 18 keV.
 Response matrices were generated using PCARSP 2.43 in
FTOOLS 5.0, with the latest energy-to-channel relationship.
        Spectral fitting was performed using XSPEC 11.0. 

 The optical counterpart of the X-ray source \source\ 
was observed with the polarimetric
detector of the High Speed Photometer (HSP) on the Hubble
Space Telescope at six different orbital phases during the
same binary orbit on 1993 August 24 12:09 UT -- August 25 
21:53 UT.
The observations were obtained in the
F277M bandpass using a 0.65 arcsec diameter aperture.  The
FWHM response of the F277M filter to a flat incident
spectrum is 2600 - 2940 \AA.  Further details on the
instrumental characteristics of the HSP can be found in
Bless et al. (1999).

       The polarization in a source was determined from its count
rate in four different analyzers,
oriented at 0$^\circ$ , 45$^\circ$ , 90$^\circ$ ,
and 135$^\circ$.  Multiple measurements in each orientation
analyzer were combined into a single set of count rates
using the method recommended by Clarke et al. (1983).  The
normalized Stokes parameters and their associated
uncertainties were then derived using the procedure outlined
by Dolan et al. (1994).   The magnitude of
the polarization and its associated uncertainty were derived
from the normalized Stokes parameters using the equations
given by Dolan \& Tapia (1986).  The magnitude of
polarization, $p$, was corrected for its non-normal
distribution at low statistical significance (Simmons \&
Stewart 1985) using the correction factor of Wardle \&
Kronberg (1974).

Flux densities in the F277M
bandpasses were calibrated by observations of BD +75$^\circ$ 325
(Bless et al. 1999), an O5p IUE spectrophotometric standard.
To derive the flux density from the polarimetric
observations, the count rates used were the average of the
summed count rates in the 0$^\circ$  and 90$^\circ$  orientation analyzers
and the summed count rates in the 45$^\circ$  and 135$^\circ$  pair (Dolan
et al. 1994).

LMC X--3 was detected in our HSP observations during one orbital
phase at the 80 $\sigma$ level of significance in the count
rate.  The relatively larger uncertainties on the derived
polarizations and the fluxes stated on an absolute scale are
caused by photometric variations resulting from small
differences in pointing after reacquisition of guide stars,
coupled with the spherically aberrated images exceeding the
diameter of the observing aperture (cf. Dolan et al 1994).

\begin{figure}[ht]
\hspace{0.5cm}
{\psfig{file=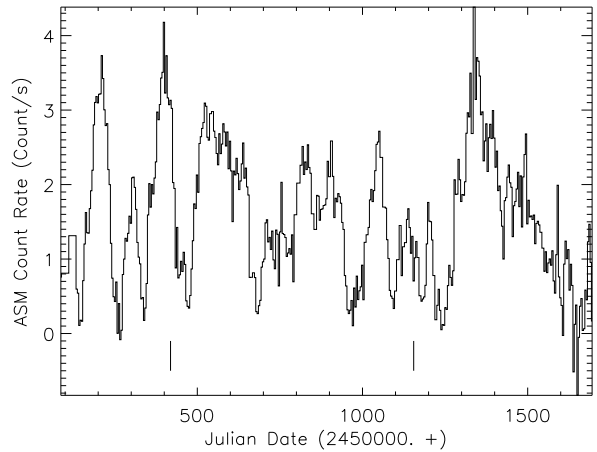,width=9cm,angle=90}}
\caption{ Long term variation of \source\ as observed by the 
RXTE All-Sky Monitor.  The four-day-average count rate is plotted
against Julian Date.  Not strictly periodic, the episodes have
time scales from 64 days to 292 days.  The PCA observations discussed
in this paper are denoted with the short vertical lines at the bottom
of the figure.  The 1998 December observations occurred when the ASM
rate was near the relatively low value of 1 count s$^{-1}$.  An archival 
observation from 1996 December,
 when the ASM rate was near 3 count s$^{-1}$, is also discussed.
}
\end{figure}

\section{Time Variability}

The long-term light curve of \source\ as measured by the All-Sky
Monitor (ASM) aboard RXTE is shown in Figure 1, where the four-day average
count rate is plotted against Julian Date.
The long term
intensity variation is apparent, and clearly not strictly periodic.
The vertical lines denote the epochs of the pointed PCA
observations discussed in this paper.  We first turn our
attention to the later of these, which began on JD 2451155,
at the moderately faint average ASM 
rate of $\sim 1$ count s$^{-1}$.
Our RXTE PCA light curve of \source\ is shown in Figure 2.  The entire 6.4
day coverage
 of our observing campaign is shown, along with 1-$\sigma$ error bars.  
The
PCA rate in counts s$^{-1}$
is plotted against time in days (upper axis) as well as running binary
phase (lower axis).  
Orbital phase is calculated using the
 van der Klis (1985) ephemeris, namely
T$_0$ = HJD 2445278.005 + 1.70479 * N.
Phase zero corresponds to superior
conjunction of the X-ray source.
The observations 
cover the $\sim3.5$ consecutive binary orbits nearly uniformly,
with only a few short gaps during this time.  

\subsection{Abrupt Flux Transitions }

It is apparent in Figure 2 that there are two relatively
flat stretches where the PCA rate stays near $\sim$75 count s$^{-1}$ for about 
one binary orbit,
each of which abruptly ends with a flux decrease of $\sim$14\% to a 
rate of $\sim$66 count s$^{-1}$.
At each
downward transition, the flux moves from the higher rate to the
lower rate over a duration of  $\leq$ 0.42 days.
The one upward and two downward transitions occur at similar binary phases.
The time at which the first downward transition
crosses the value of 70 count s$^{-1}$
is  $\sim$MJD 51158.15,  phase
0.19.  The time at which the upward transition crosses this level (on its
return to the "nominal" flux level)  is  $\sim$MJD 51159.85,
again binary phase 0.19---very nearly one integer binary orbit later.
While the coverage gap immediately preceding the second downward transition
makes the measured onset time uncertain  (see Figure 2),
we estimate it
begins at or before $\sim$MJD 51161.74,
phase $\leq$ 0.29.  In other words, the cycle from one downward
transition to the next repeats at very nearly
\emphasize{twice} the orbital period.  This is interesting, for it implies
that the
mechanism giving rise to the abrupt flux transitions is dynamically linked
to the orbit.



\begin{figure}[ht]
\hspace{0.5cm}
{\psfig{file=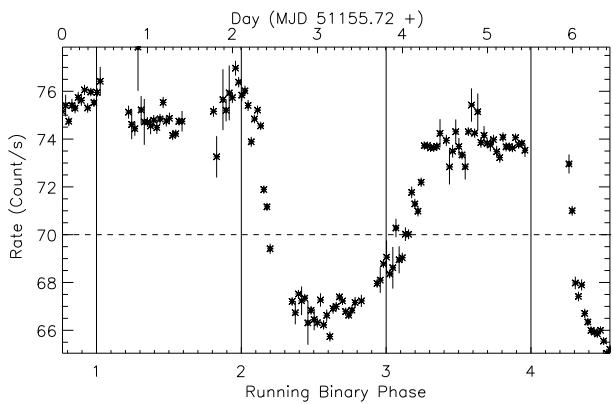,width=9cm,angle=90}}
\caption{PCA
 light curve of \source\ versus phase (lower X-axis) and
 time in MJD (upper X-axis) from 1998 December.
 The count rate is for a total of 3 PCUs.  The time resolution is
 3200 seconds. Vertical lines separate consecutive binary orbits.
The horizontal dashed line represents our threshold for folding data
at high and low flux (see text).  An overall gross flux
 variation of $\sim$14\% is evident, with abrupt flux transitions
recurring at $\sim$twice
 the orbital period of the binary.  Less obvious at this scale is the
sinusoidal \emphasize{orbital} modulation, with an amplitude $\sim$5\%.
}
\end{figure}
\subsection{Orbital Phase Dependence of X-ray Flux}


We searched for evidence of X-ray modulation at the orbital period by
dividing the data into two groups based on source counting
rate. Data with a rate below 70 c/s were folded on the orbital period
separately from those points with rate above 70 c/s, thus preventing the
gross flux transitions discussed above from contaminating the
folded light curves. (Taking the data set as a whole and folding on the
orbital period would result in data from higher and lower gross flux
segments being averaged together in a single phase bin, which masks
the lower-amplitude phase modulation.)



Figure 3 shows the results of folding each of these data sets on the
orbital period of the van der Klis (1985) ephemeris.
The orbital modulation is clear in the low-flux data (Fig. 3a)---it
reaches a maximum slightly beyond $\phi=0.0$ while the minimum occurs
near $\phi=0.5$.  
Folding the 
high-flux data (Fig. 3b) at the orbital period shows apparently weaker
phase modulation.

To further investigate the presence of phase dependent flux variability, we
performed periodogram analysis on the data set as a whole, binned
at 800 s,
and the phase-selected subsets defined above.  We restricted the
period search to within 0.5 days of the observed optical orbital
period.  Significance was
estimated using the the false-alarm
probability defined by Scargle (1982). 
The
best-fit trial period of the low-flux data is 1.693$\pm$0.037 d,
with a false-alarm probability of 2.4$\times$10$^{-14}$ for a $>$ 7$\sigma$
detection.  
The best-fit trial period
for the high-flux data is 1.712$\pm$0.037 d, with false alarm
probability 
of 4.5$\times$10$^{-11}$ for a $>$ 6$\sigma$ detection.  Each detection
is significant and consistent with the observed orbital period in the
optical.
We conclude that significant
periodic amplitude modulation 
is present in the X-ray flux from LMC X--3, and that
the amplitude of the modulation
is slightly higher when the 2-10 keV flux is lower.
 This is the first reported detection of orbital phase
 dependence in the X-ray intensity from LMC X--3.

\begin{figure}[ht]
\hspace{0.5cm}

{\psfig{file=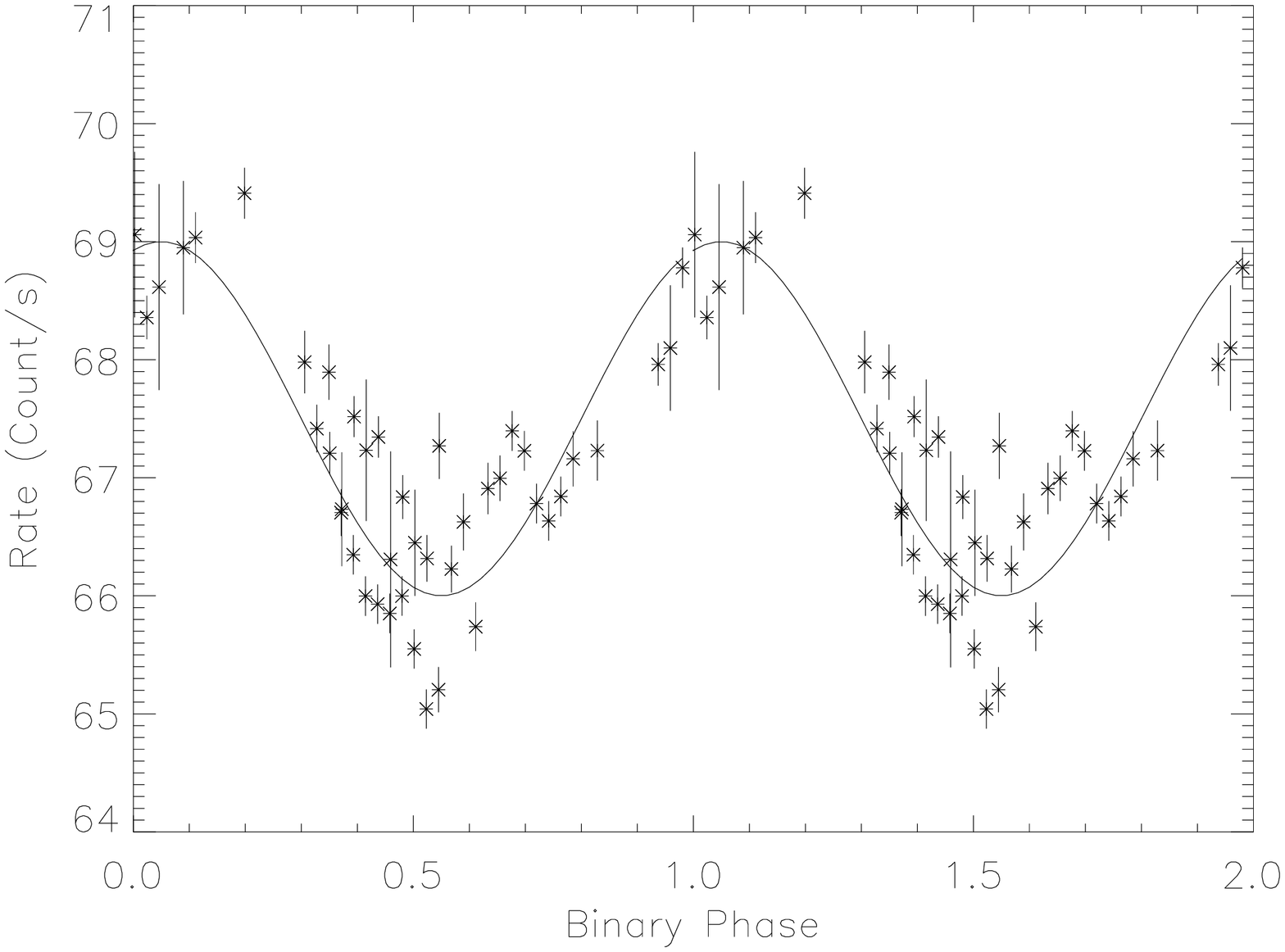,width=9cm,angle=90}}
{\psfig{file=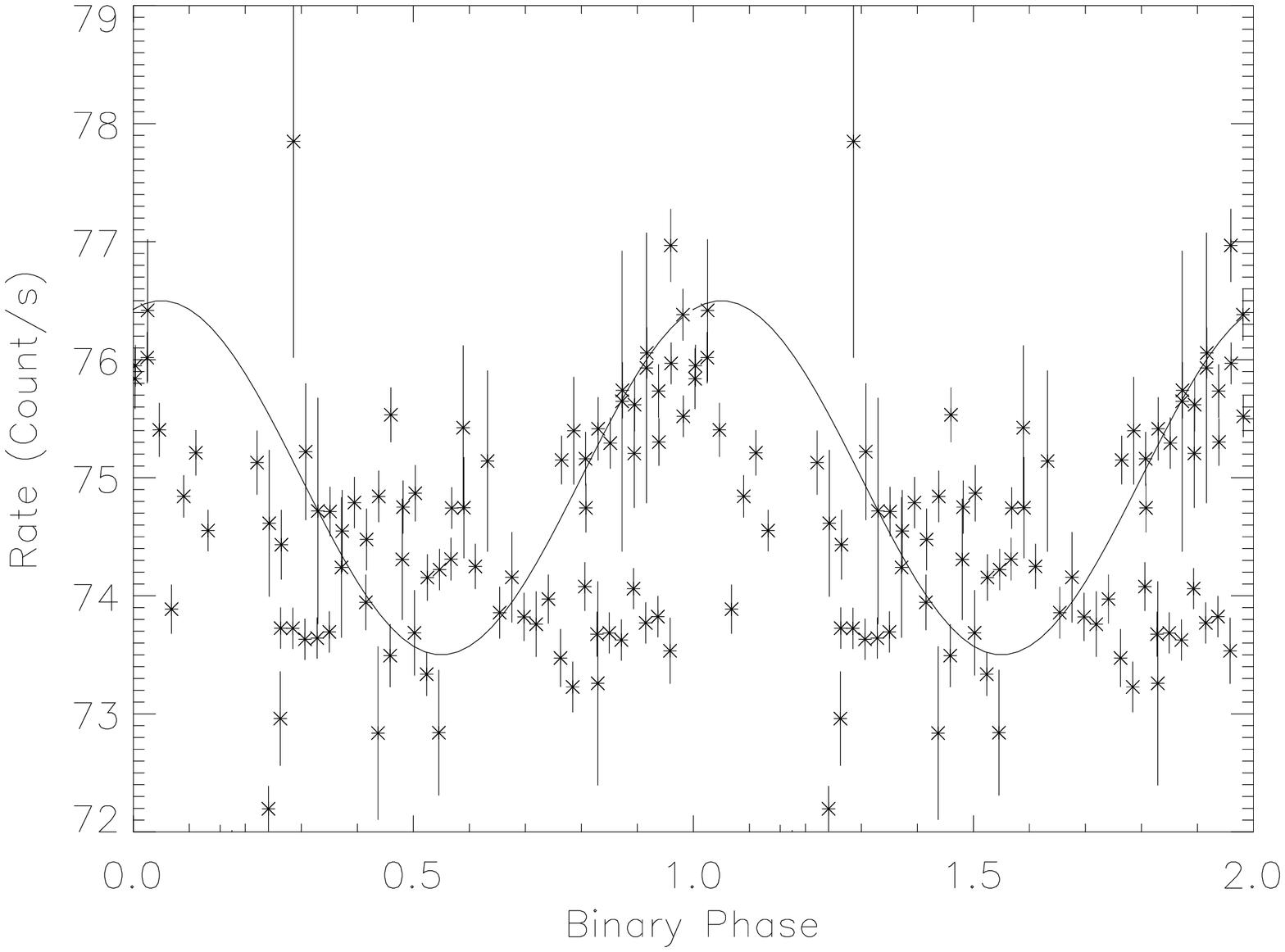,width=9cm,angle=90}}

\caption{
The PCA light curve from the 1998 December observations, folded on the
1.7 day binary period.  The top panel includes only data where the 
rate is $<$ 70 count s$^{-1}$; at these lower fluxes, the amplitude of the orbital
variation is 7\%.  The lower panel shows the floded higher-flux data
($>$ 70 count s$^{-1}$). Phase variability at this count rate is less significant.
A sine wave with the same period and phase
has been superimposed on each plot to guide the eye.
}
\end{figure}

\subsection{Archival RXTE Observations during a high/soft state}

Nowak et al. (2000) discuss a previous 140 ks RXTE observation
of LMC X--3 obtained on
1996 November 30 -- December 6, when the source was at the relatively high 
RXTE/ASM count
rate of about 3.5 count s$^{-1}$.  Their light curve (Figure 1 of
Nowak et al. 2000) shows
similar qualitative behavior: a slow flux transition from higher rate to lower
rate in the 2-10 keV flux with
an overall amplitude $\sim12\%$.  Motivated by the detection of strong
phase dependence in \source\ at relatively low flux levels, we extracted
this archival data set to search for orbital modulation at a significantly
different flux level and epoch.  Since different detectors were on/off 
through this observation, we scaled the lightcurve to the 5-pcu rate.
Again, due to the overall gross flux transition
present during this long observation, we grouped their data into two segments,
selected by total PCA rate (above/below 410 count s$^{-1}$), 
before folding on the van der Klis (1985) ephemeris.  The
results of this analysis are presented in Figure 4.  Phase dependence is
clearly present in both groups of data.  The phase coverage of the observation
when the source was at slightly greater flux levels is insufficient to
trace out an entire cycle; only orbital phase $\phi$ = 
0.0-0.5 was observed.  At
the lower count rate, the phase coverage is sufficient to trace out a
light curve qualitatively similar to that seen in the low flux
1998 December data.
As in the 1998 December observation,
it is single-peaked;
the maximum X-ray intensity occurs near $\phi=0.95-0.05$
 while
the minimum is near $\phi=0.5$ .  As in the 1998 December data, the amplitude
of this modulation is $\sim5\%$.

\begin{figure}[ht]
\hspace{0.5cm}

{\psfig{file=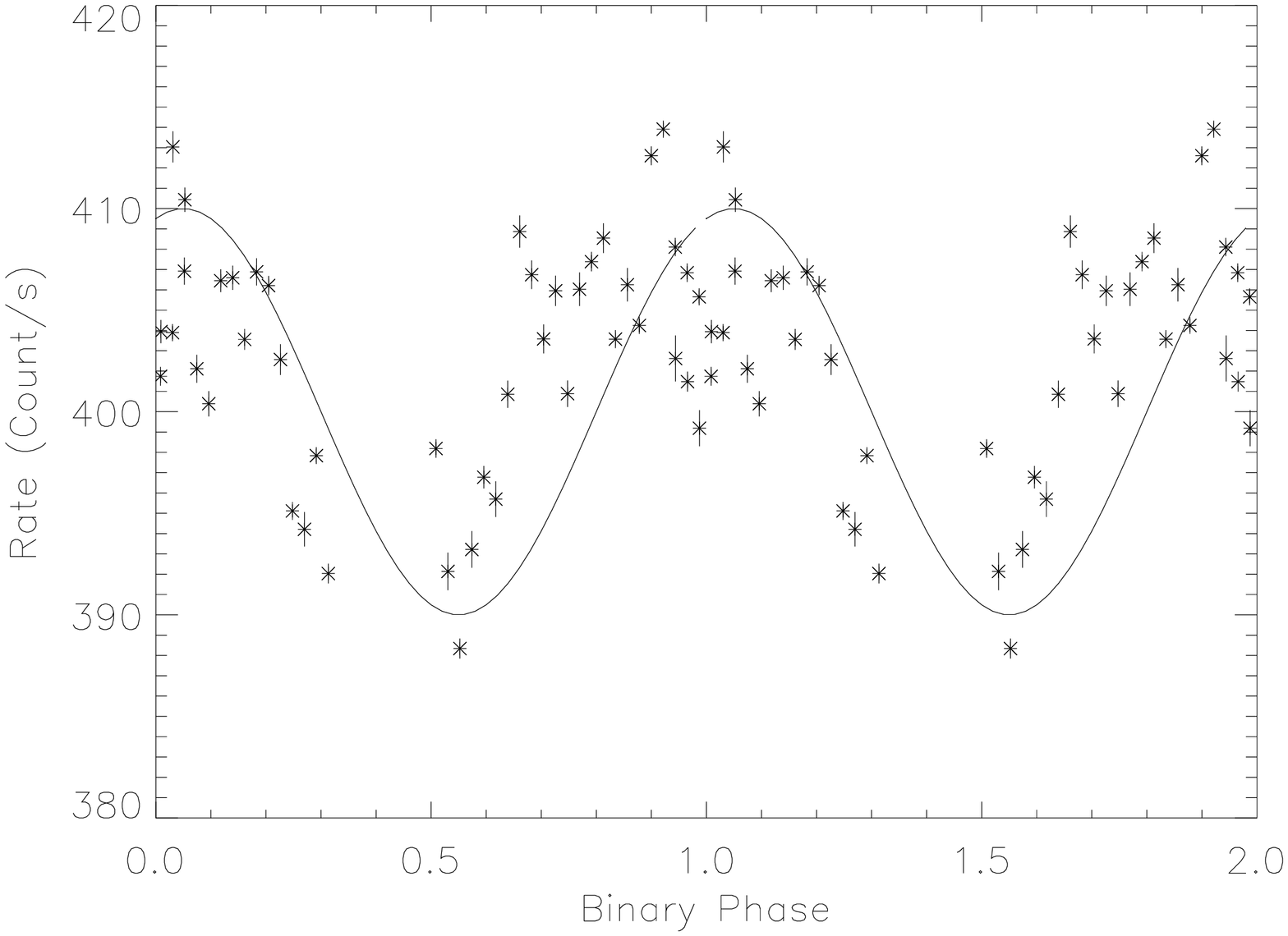,width=9cm,angle=90}}
{\psfig{file=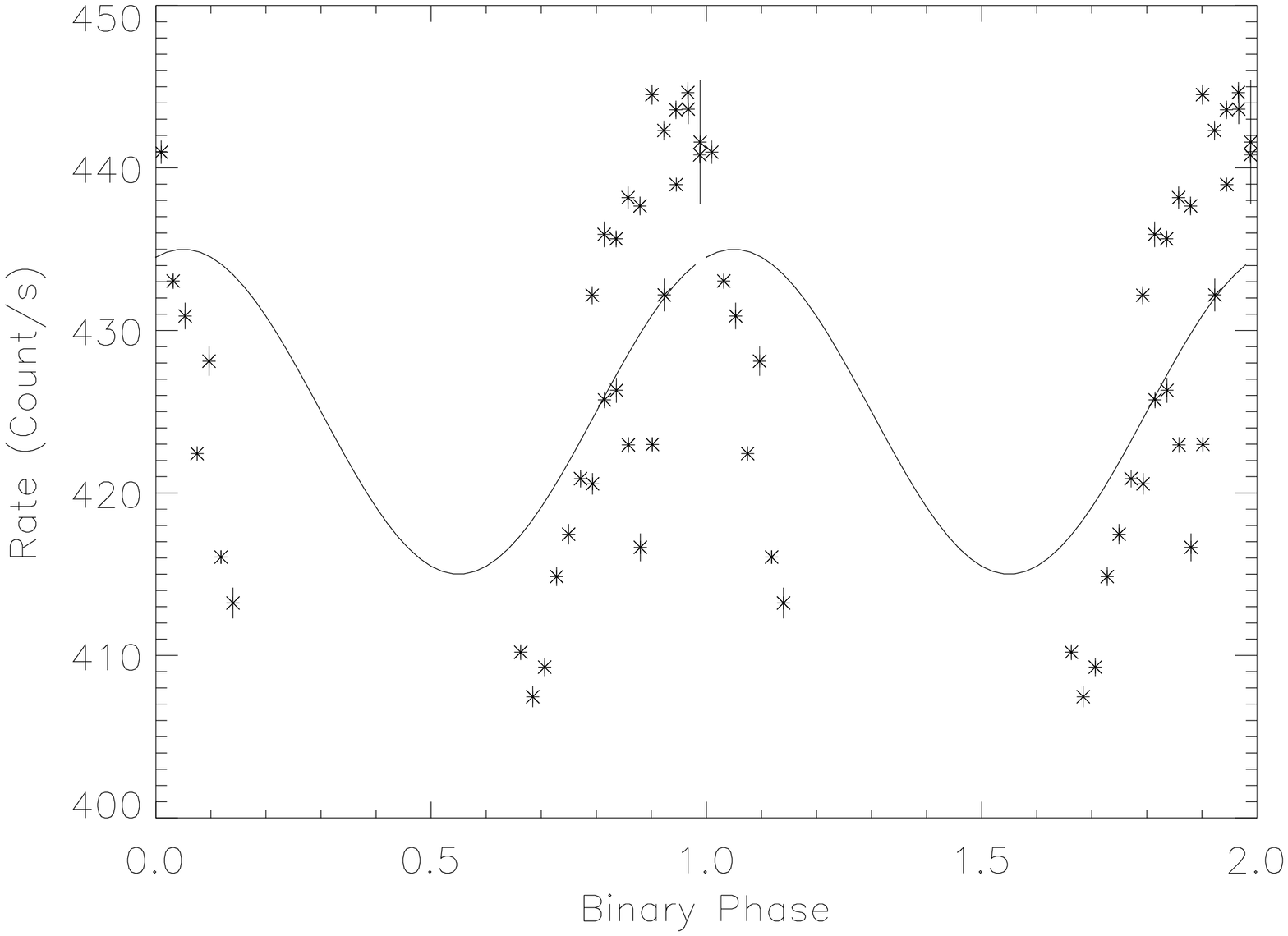,width=9cm,angle=90}}

\caption{
Archival PCA observation of LMC X--3 from 1996 December,
 folded on binary phase.  Rate is for 5 PCUs.
(See Nowak et al. 2000, Figure 1, for discussion.)
Top panel: data below 410 count s$^{-1}$.
Bottom panel: data above count s$^{-1}$.
The ASM rate during this observation was about 3.5 count s$^{-1}$.
The phase-folded data 
show a light curve shape similar
to the 1998 December observation,
when the ASM rate was 1 count s$^{-1}$.  A sine wave with the same phase and period
as shown in the previous figure has been superimposed on both panels to
guide the eye.
}
\end{figure}

\subsection{Searches for Rapid Time Variability}

We searched for rapid aperiodic variability in \source\ using the high
time resolution PCA data obtained in the Good Xenon mode.  Discrete
power spectral density distributions (PSDs) were calculated by
dividing the data into segments of uniform length, performing fast
Fourier transforms of each, and averaging the results.  The PSDs were
normalized such that their integral gives the squared RMS fractional
variability (Miyamoto et al 1991; van der Klis 1989). We subtracted
the Poisson noise level from the power spectra, taking into account the
modifications expected from PCA detector deadtime.
In its low/hard state, LMC X-3 displays quasi-periodic oscillations
(QPO) at a centroid frequency of 0.4 Hz, and significant time
variability of 30--50\% (Boyd et al. 2000). Neither are detected in
this observation, and the long duration of this data set allows us to
place a rather stringent upper limit of 1\% on the rms amplitude of a
QPO feature between 0.1--10~Hz.

\subsection{Summary of Timing Results}

Our analysis of the time variability in LMC X-3 indicates that the light
curve shown in Figure 2 can be described as the superposition of 1) an
orbital modulation of roughly constant amplitude and 2) abrupt flux
transitions recurring at twice the orbital period.  Significant orbital
modulation is also present in archival data (Figure 4) taken when LMC X-3
was at a dramatically different flux level, implying that orbital modulation
is a reliable feature in the X-ray output through a fourfold range of source
intensities.  The 1996 December archival observation also contains one gross
flux transition, though the data set is not long enough to determine whether
it repeats, and if so, at what period.  The amplitude of the orbital modulation
in the 1996 December observation is approximately 5\%, as in the 1998
December data.

\section{HSP Results: UV Photometry and Polarimetry}

	Orbital modulation of the X-ray flux has not been
previously observed in \source, although it is a well-known
feature of the optical light curve.  In the V-band, the ellipsoidal
deformation of the B-star results in the double-peaked
morphology of the light curve.  In an effort to extend our
knowledge of the orbital modulation of \source\ as a function of
photon energy, we present the results of an ultraviolet
photometric and polarimetric
study of \source\ using the High Speed Photometer aboard the Hubble
Space Telescope.

\begin{figure}[ht]
\hspace{0.5cm}
{\psfig{file=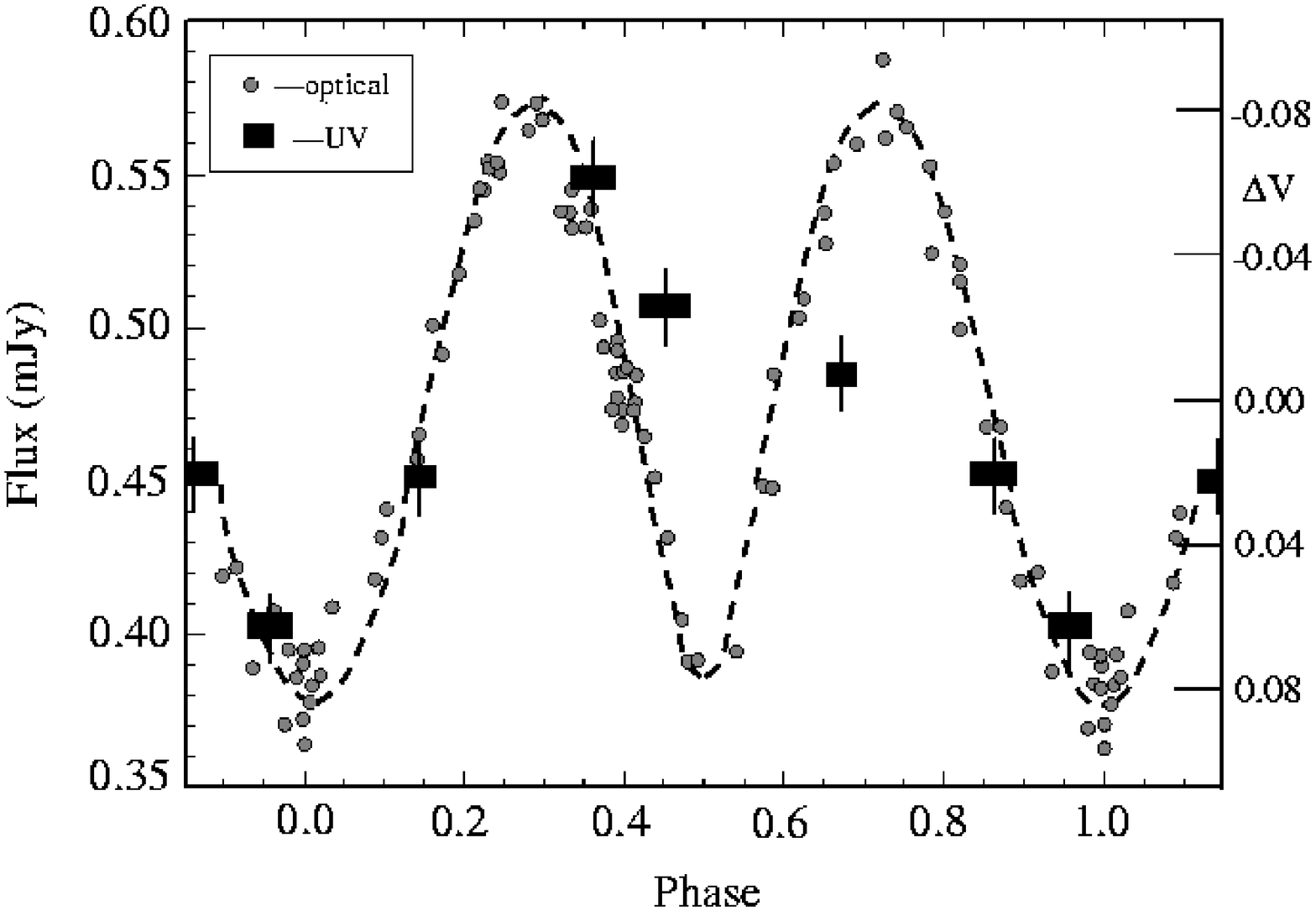,width=9cm,angle=270}}
\caption{The UV lightcurve from HST obtained in 1993 (this paper, rectangles) 
superimposed on the optical V lightcurve from
1984 (van Paradijs, 1987, circles).
UV flux is given on the left hand axis; V-band magnitude on the right
hand axis.  The two scales are independent; their intervals were
adjusted to give the same amplitude in each band pass.
While its phase coverage 
is sparse, the UV variability is consistent with
the optical light curve.  
}
\end{figure}

In Figure 5 we compare the
orbital light curve we obtained in the F277M bandpass in
1993 with the V band light curve obtained in
1984 by van Paradijs et al. (1987).
While the phase coverage of the UV light curve is sparse,
the overall variation is not inconsistent with the double-peaked
optical light curve.  
The UV light curve 
is also consistent with 
an additional source of UV radiation in the system, with maximum
occurring near
inferior conjunction of the X-ray source.  This second light would
presumably be
coming from the accretion disk.

No linear polarization was detected at any orbital phase
(see Table 2).  Assuming constant polarization from LMC X--3
in the F277M bandpass, and treating the normalized Stokes
parameters as normally distributed (but cf. Clarke et al.
1983), the 2 $\sigma$ upper limit on the linear polarization
of LMC X--3 in the F277M bandpass is p $\leq$ 0.018.
Several other X-ray binary systems, notably Cygnus X-1, do show
significant phase-dependent polarization in the
UV, interpreted as evidence for
single scattering off 
gas streams connecting the primary to the accretion disk (Dolan \& Tapia
1988, Wolinski et al. 1996). The upper limit on 
any variable polarization is too large to contain any information
about the structure of the gas streams in the system.

\section{X-ray Spectral Variability}

	There is no single accepted model for the spectral variability of
galactic black hole candidates across their several spectral states 
(see reviews by Tanaka \& Lewin 1995, Nowak 1995, van der Klis 1995, 
 for discussions of black hole spectral
states).  Numerous authors choose a phenomenological "disk black body
plus power law" model, in which the spectrum is fit by the 
superposition of multiple rings of black bodies representing
the cooler accretion disk, plus a power law to describe the hard photons.
While not motivated by a clear physical picture of what gives rise to
these independent components, this model does give reasonable results for
spectral fits (though commonly a "feature" is seen at the cross-over
point of the two functions).  Alternative models based on a physical
description of the interaction of a spherical Comptonizing corona and
cooler seed photons from the disk (Shrader \& Titarchuk 1999, Borozdin
et al., 1999) are often only applicable in restricted spectral states,
or produce biased residuals at high energies.  Because previous spectral
studies of \source\ have concentrated on the phenomenological
"disk black body plus power law" model (Wilms et al. 2000, Nowak et al.
2000, also Ebisawa, Cowley), and because of the 
ambiguity in the spectral state of \source\
at this relatively low flux level, we have chosen to report our
spectral fitting results
using this model.  In the following discussion, model parameters for
the powerlaw component are photon index
$\Gamma$ and $A_{pl}$ in photons keV$^{-1}$ cm$^{-2}$ s$^{-1}$.  The disk black
body model is parameterized by $T_{in}$, the temperature (kT) at the inner
edge of the accretion disk, as well as a geometric factor $N$ which 
depends on the inner disk radius, the viewing angle of the accretion
disk, and the distance to the source.

Wilms et al. (2000) report recurrent low/hard
states in LMC X--3 based on long-term monitoring with RXTE 
over several years.  This was confirmed by more recent RXTE observations
during a low/hard state (Boyd
et al. 2000) in which both the timing and spectral characteristics of
black hole candidates in the low/hard state was observed.
Wilms et al. (2000) report a disk temperature that
drops below 1.0 keV as the source enters the low state.
The source is well described by a pure power law with $\Gamma=1.7$
when truly in the low
state.	

	In our investigation of the spectral changes during our
observation, we restrict ourselves to 
data obtained far from SAA passage and with low measured electron
contamination.  We only consider observations where the orbital phase
was 
between 0.475 to 0.525
in an attempt to minimize orbital effects when
comparing the higher flux and lower flux spectra.
Our spectral fits include
data from 2.5-25.0 keV.  The 
power law plus black-body disk model was found
to be acceptable under the $\chi^{2}$ test for each individual spectrum,
with typical reduced $\chi^{2}$ values near or below 1.0 for 49 degrees
of freedom.    
	
Results of spectral fits to the $\phi=0.5$ observations are summarized
in Table 3.  A typical two-component model fit is shown in Figure 6. 
The disk temperature hovers near the critical value of $kT=1.0$
reported by Wilms et al. (2000) as an indicator of the occurrence of
a transition from the high/soft to the low/hard state.  Unlike
the pure low/hard state however, a significant
disk component is required to achieve a reasonable spectral fit to the
current observations.
This may indicate that \source\ was undergoing a state transition at
the time.  The model flux reported in Table 3 is about 15 times greater
than the flux measured during the 2000 May low/hard state (Boyd et al. 2000).

The
disk black body spectral components show a roughly linear trend with
the 2-10 keV flux.  The changes in $T_{in}$ are small, and within the
formal uncertainties on this parameter.  On the other hand, the disk
normalization parameter $N$
varies by $\sim 20\%$ from minimum 
to maximum flux.  Since $N$ is proportional to $r^{2}_{in}$cos$\theta$
this implies either a change in the inner accretion disk radius, or
the inclination of the disk, is responsible for the gross flux variation
from one cycle to the next.  A global warp or kink, traveling around the
disk, could cause the effective inclination of the disk to vary from one binary
cycle to the next.  

Nowak et al. (2000) point out that, during the long
observation in 1996 December, variations in the disk black body
spectral components run counter to previously observed long term trends,
wherein the normalization term remains fairly constant while the disk
temperature exhibits marked variability with observed flux.  Our
results of the 1998 December observations follow the 
short time scale trend found by Nowak et al. (2000) where the disk
black body temperature is relatively constrained and variability is
instead correlated to the geometric normalization parameter.

\begin{figure}[ht]
\hspace{0.5cm}
{\psfig{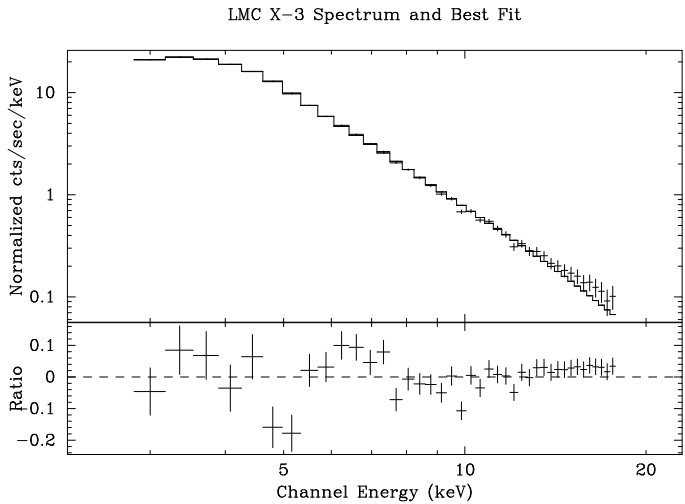}}
\caption{PCA spectrum of LMC X--3 and best-fit disk black body plus
power law model for observation 24 (in Table 1).  
The ratio of the model to the data is shown in the lower
panel.   Parameters of the fit are listed in Table 3.
}
\end{figure}

\section{Discussion and Conclusions}

Our RXTE observations of \source\ through 3.5 consecutive binary cycles
have detected significant orbital phase modulation of the
2-10 keV X-ray flux.  
The amplitude of the orbital modulation is variable,
being  higher when the the system experienced an abrupt decrease
in flux.  Comparison of our
results to an archival observation at a profoundly
different flux level shows that phase-dependent flux is a hallmark of
\source\ through a wide range of overall X-ray intensities, with an
amplitude of between 5--10\%.
The UV light curve, while sparse, is
consistent in shape with the optical variability, implying that
the dominant component of the UV radiation is the deformed main
sequence star and outer disk, as in the optical.

A phase-dependent X-ray flux may be due to a) a compact "hot spot" of
X-ray generation moving into and out of our line of sight, b) a region
of cooler absorbing matter moving into and out of our line of sight,
or c) a combination of the two.  Since \source\ does not display
eclipses of the compact object, its inclination is not well constrained.
A possible geometric model that explains the variation seen at the
different flux levels is that of
 a hot spot located between the main sequence star
and the compact object, 
perhaps where the stream impacts the 
accretion disk.
 X-rays from this  
hot spot could be absorbed by the accretion disk at
inferior conjunction of the X-ray source
($\phi = 0.5$) but be visible to the observer at superior conjunction
($\phi = 0$).
The expected curvature of the gas stream would give rise to the flux
maximum occurring slightly after superior conjunction.  Changes in the
location of the gas stream as a function of mass accretion rate could
account for the slightly
different phase of maximum during the high state 1996
observations.

An equally striking feature of our RXTE light curve is the abrupt flux
decreases, separated by very nearly two binary periods.  The slower return to
the higher flux level also occurs near similar binary phase.
One possible model for the system that explains both the overall flux
transitions and the variable orbital phase modulation is that of a
disk excited into a low-order global mode.  Global
instabilities in highly ionized accretion disks around compact objects
are not sufficiently well understood to claim that this model is
realistic or expected.  However, there is growing numerical and
observational evidence that suggests similar global structures
in accretion disks do develop and persist.   Such a large-scale perturbation
of the accretion disk provides a global mechanism for angular momentum
transport and dispersion which, together with the local viscous
processes, can drive the dynamics of the disk.
Evidence for significant two-armed spiral structure has been seen in
Doppler tomography studies of 
the dwarf nova IP Pegasi during two outbursts (Steeghs, 2000).   
Three dimensional hydrodynamical simulations of accretion disks 
show that the formation of spiral shocks
is common over a wide range of system parameters, including disk
temperatures much hotter than those in dwarf novae (Boffin et al. 1999).
We suggest the presence of a global density wave in the accretion
disk of \source\
as one possible explanation that
is consistent with the data.
If such a global feature were stable
for several binary orbits, and rotated with an angular velocity close
to one quarter the orbital velocity, then the gross features in the
X-ray light curve would be reproduced due to the varying obscuration
along the line of sight.  As the feature rotated around the disk it
would at times obscure observed X-rays in a phase-dependent fashion,
and at other times have little appreciable effect on the observed
X-ray flux.  This concept is illustrated schematically in Figure 7,
where we represent the global mode as a spiral density wave. 
We do not rule out other combinations of disk-- and orbital--geometry
that may explain the observed behavior.  We note, however, that the
spiral structure seen in IP Peg, as well as large scale disc asymmetries
in other systems, have been observed to co-rotate with the binary
system.

 \begin{figure}[ht]
 \hspace{0.5cm}

 {\psfig{file=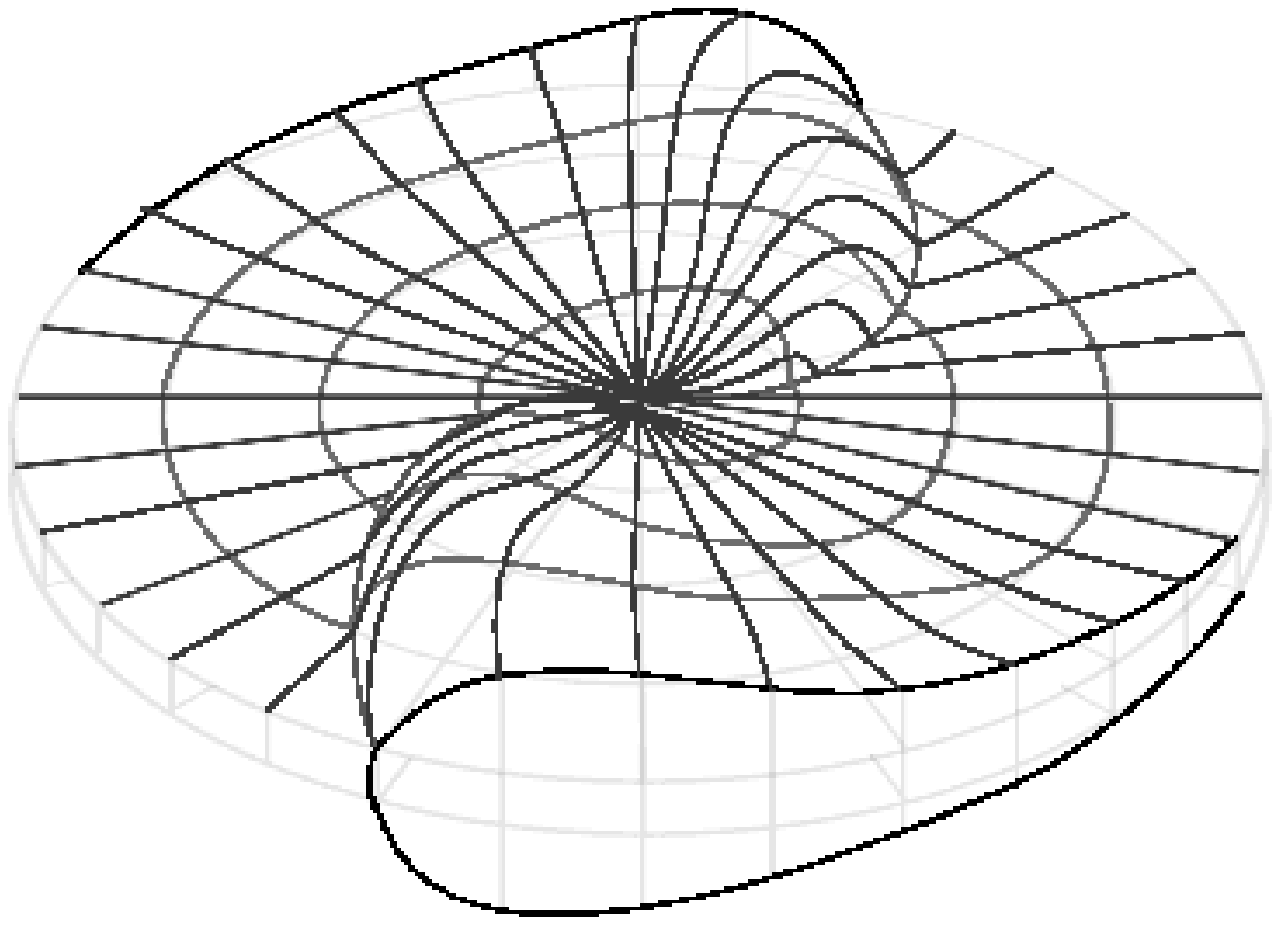,height=6cm,width=9cm}}

 \caption{
 A schematic representation of a low order global oscillation mode
 excited in an accretion disk.  If such a feature were stable
 for several binary orbits, and rotated with an angular velocity close
 to one quarter the orbital velocity, then the gross features in the
 X-ray light curve would be reproduced due to the varying obscuration
 along the line of sight.
 }
 \end{figure}

The long term variability of \source\ is not easily characterized.  It
is becoming clear that the previous conjecture of periodic large amplitude
flux variations arising from a warped precessing disk is not complete:
the large scale variation seen in the RXTE ASM (Figure 1)
 is far from simply periodic.
If instead its long term behavior is similar to Cygnus X-1 and
other persistent galactic black hole candidates, then \source\ stands out
due to the suggestion of a preferred time scale: on the order of hundreds
of days from one low/hard state to the next.  If a global disk instability is
responsible for state transitions in such systems then the disk dynamics in
\source\ must be quite different from the persistent galactic black holes,
which apparently undergo the transition in a more random fashion.
It is possible that \source\
does not fit neatly into either category above, but has characteristics
of each.  In that case, continued intensive studies of the system should
be carried out with the goal of developing a consistent physical model of
the accretion disk structure and dynamics.

\acknowledgments

This research has made use of data obtained through the High Energy
Astrophysics Science Archive Research Center Online Service, provided
by NASA's Goddard Space Flight Center. ASM results were provided
by the ASM/RXTE teams at MIT and at the RXTE SOF and GOF at NASA's
GSFC.
Based in part on observations with the Hubble Space Telescope
obtained at the Space Telescope Science Institute, which is 
operated by AURA, Inc. under NASA contract NAS5-26555.
We enjoyed helpful
conversations with Mike Nowak and Joern Wilms.  Karen Smale produced
the artwork in Figure 7.

\clearpage

\begin{deluxetable}{lllll}
\tablewidth{0pc}
\tablecaption{LMC X--3 RXTE Observation Log, 1998 Dec 8 - 15}
\tablehead{
\colhead{Obs} &
\colhead{Start MJD} &
\colhead{Exp Time (s)} &
\colhead{Orbital Phase} &
\colhead{Rate (count s$^{-1}$} 
}
\startdata
       1 &      51155.719  &     17032.0 & 0.765 & 74.30 \nl
       2 &      51156.042  &     10592.0 & 0.955 & 75.77 \nl
       3 &      51156.516  &     2762.0 & 0.232 & 74.36 \nl
       4 &      51156.633  &     1216.0 & 0.301 & 75.89 \nl
       5 &      51156.700  &     14248.0 & 0.340 & 74.95 \nl
       6 &      51156.993  &     11208.0 & 0.512 & 74.10 \nl
       7 &      51157.512  &     1920.0 & 0.816 & 75.04 \nl
       8 &      51157.633  &     1368.0 & 0.887 & 76.71 \nl
       9 &      51157.701  &     16404.0 & 0.929 & 76.18 \nl
       10 &     51158.000  &     15355.0 & 0.103 & 74.01 \nl
       11 &      51158.422 &      12740.0 & 0.350 & 68.18  \nl
       12 &      51158.710 &      14058.0 & 0.520 & 67.62 \nl
       13 &      51158.970 &      18031.0 & 0.671  & 67.89 \nl
       14 &      51159.242 &      1340.0 & 0.831 & 69.08 \nl
       15 &     51159.421  &     11099.0  & 0.937 & 69.47 \nl
       16 &     51159.710  &     14266.0 & 0.106 & 71.14 \nl
       17 &     51159.969  &     16895.0 & 0.258 & 74.25 \nl
       18 &     51160.241  &     1664.0 & 0.418 & 75.51 \nl
       19 &     51160.309  &     1536.0 & 0.457 & 76.03 \nl
       20 &     51160.381  &     14087.0 & 0.500 & 74.86 \nl
       21 &     51160.710  &     14749.0 & 0.693 & 73.89 \nl
       22 &     51160.967  &     17920.0 & 0.844 & 74.36 \nl
       23 &     51161.709  &     22855.0 & 0.280 & 67.89 \nl
       24 &     51162.042  &     12544.0 & 0.474 & 66.54 \nl

\enddata
\end{deluxetable}

\begin{deluxetable}{lllll}
\tablewidth{0pc}
\tablecaption{UV Photometry and Polarimetry of LMC X--3, 1993 August 24 - 25}
\tablehead{
\colhead{Orbital Phase} &
\colhead{Flux (mJy)} &
\colhead{Polarization} &
}

\startdata

0.655 - 0.687 &    0.485 $\pm$ 0.012 &     0.016 $\pm$ 0.018 \nl
0.835 - 0.891 &    0.452 $\pm$ 0.012 &     0.014 $\pm$ 0.024 \nl
0.934 - 0.981 &    0.402 $\pm$ 0.011 &     0.000 $\pm$ 0.030 \nl
0.127 - 0.161 &    0.451 $\pm$ 0.012 &     0.000 $\pm$ 0.022 \nl
0.334 - 0.389 &    0.549 $\pm$ 0.013 &     0.000 $\pm$ 0.021 \nl
0.423 - 0.480 &    0.506 $\pm$ 0.012 &     0.017 $\pm$ 0.017 \nl

\enddata
\end{deluxetable}

\begin{deluxetable}{lllllll}
\tablewidth{0pc}
\tablecaption{Parameters of $\phi=0.5$  Spectral Fits}
\tablehead{
\colhead{$L_{x}$ x 10$^{37}$ erg s$^{-1}$} &
\colhead{Orbital Phase} &
\colhead{$\Gamma$} &
\colhead{$A_{pl}$ } &
\colhead{$T_{in}$ (keV)} &
\colhead{$N$} &
\colhead{Obs } 
}

\startdata

10.96 & 0.512 & $3.24^{+0.06}_{-0.40}$ & $0.16^{+0.021}_{-0.103}$ & $1.01^{+0.010}_{-0.004}$ &  $27.5^{+2.60}_{-0.50}$ &   6 \nl
10.72 & 0.500 & $2.83^{+0.01}_{-0.16}$ & $0.087^{+0.01}_{-0.015}$ & $1.02^{+0.001}_{-0.013}$ &  $27.2^{+0.50}_{-0.50}$ &   20 \nl
9.63 & 0.52 & $3.03^{+0.07}_{-0.13}$ & $0.21^{+0.039}_{-0.056}$ & $0.98^{+0.008}_{-0.012}$ &  $21.9^{+2.00}_{-1.90}$ &   12 \nl
9.39 & 0.474 & $3.08^{+0.07}_{-0.13}$ & $0.21^{+0.069}_{-0.053}$ & $0.954^{+0.015}_{-0.002}$ &  $23.9^{+1.50}_{-1.10}$ &   24 \nl

\enddata
\end{deluxetable}

\end{document}